# Transformation equations for the kinetic energy of tardyons and photons: Filling the gap in special relativity literature

**Bernhard Rothenstein,**
"Politehnica" University of Timişoara,
Physics Department, Timişoara, Romania
**Doru Păunescu**
"Politehnica" University of Timişoara,
Department of Mathematics, Timişoara Romania

*Abstract. Transformation equations for the kinetic energy of a tardyon are derived in the limits of classical and of special relativity theory. Two formulas are presented. In the first one the energy of the particle in one of the involved reference frames is presented as a function of its proper energy, of the relative velocity of the two frames and of its speed in the second one. In the second one the kinetic energy in one of the involved reference frames is expressed as a function of its kinetic energy in the second one of its proper energy, of the relative velocity of the involved inertial reference frames and of its velocity relative to that frame. The obtained results are extended to the case of a photon that moves under the same geometrical conditions, recovering the formulas that account for the relativistic Doppler Effect, illustrating the behavior of a transformation equation when it should account for the properties of an electron and for those of a photon as well.*

The dynamic properties of a free particle (tardyon) are characterized in classical relativity by its mass, momentum and kinetic energy. The involved inertial reference frames where from the particle is detected are $I(XOY)$ and $I'(X'O'Y')$ respectively The corresponding axes of the two reference frames are parallel to each other. The $OX(O'X')$ axes are overlapped. $I'$ moves relative to $I$ with constant speed $V$ in the positive direction of the overlapped $OX(O'X')$ axes. At the origin of time in the two frames ($t=t'=0$) the origins O and O' are located at the same point in space. In the limits of special relativity theory the physical quantities which define the dynamic properties of the free tardyon are its rest mass, rest energy, momentum, energy and kinetic energy. Textbooks present transformation equations for relativistic momentum and energy but avoid to present transformation equations for the kinetic energy probably because it is not a component of a four vector.[1,2,3] The purpose of our paper is to feel that gap.

Making the convention to present physical quantities measured in I as primed and those measured in I' as unprimed, the work-kinetic energy reads in I in the limits of classical mechanics

$$dW = F_x dx = F_x u_x dt \tag{1}$$

whereas in I' it reads

$$dW' = F'_x dx' = F'_x u'_x dt' \tag{2}$$

in accordance with the fact that the laws of mechanics are the same in all inertial reference frames in relative motion. In (1) and (2) (*W,W'*), (*F,F'_x*), ($u_x, u'_x$) and (*t,t'*) stand for the kinetic energy, the OX(O'X') components of the force acting upon the particle and time. The forces and the speed have only components along the overlapped axes. Combining (1) and (2) the result is

$$\frac{dW}{dW'} = \frac{F_x}{F'_x} \frac{u_x}{u'_x} \frac{dt}{dt'}. \tag{3}$$

In the limits of classical mechanics and classical relativity mass (*m=m'*), force ($F_x = F'_x$) and time intervals (*dt=dt'*) have the same magnitude when measured from different inertial reference frames. The speeds *u* and *u'* are related by

$$u = u' + V \tag{4}$$

changes in speed having the same magnitude, when measured from I and I' respectively. Taking all that into account (3) becomes

$$dW = dW'\left(1 + \frac{V}{u'_x}\right) = dW' + Vm'du'_x \tag{5}$$

leading after integration to

$$W = W' + Vm'\Delta u'_x. \tag{6}$$

in accordance with the result obtained in[4] from other considerations.

In special relativity theory the dynamic properties of a particle are characterized in I by its rest mass $m_0$ momentum $p(p_x,p_y)$, rest energy $E_0$, energy *E* and kinetic energy *W*. The corresponding physical quantities are when detected from I' $m_0$, momentum *p'*($p'_x, p'_y$), rest energy $E_0$, energy *E'* and kinetic energy *W'*, The particle moves with sped $u(u_x,u_y)$ relative to I and with speed *u'*($u'_x, u'_y$) relative to I' the magnitudes *u* and *u'* transforming as

$$u = u' \frac{\sqrt{(\cos\theta' + \frac{V}{u'})^2 + (1 - V^2/c^2)\sin^2\theta'}}{1 + \frac{Vu'}{c^2}\cos\theta'} \tag{7}$$

*θ'* representing the angle made by direction in which the particle moves and the positive direction of the overlapped axes.[5]

The kinetic energy of a particle is a measure for the mechanical work it could perform when we reduce its speed from the initial magnitude to zero. Bertozzi's experiment[6] is a landmark in our understanding of relativistic dynamics convincing us that the concept of absolute mass does not explain the results of experiment in which particles are accelerated up to speeds comparable with the speed c at which light signals (photons) propagate in empty space. Considering that Bertozzi's experiment is performed in the I' inertial reference frame the involved electrons are accelerated in the positive direction of the overlapped OX(O'X') axes then let *W'* be their kinetic energy. Detecting the same experiment from I the kinetic energy of the same particle is *W*. The problem is to find out a relationship between *W* and *W'* i.e. to derive a transformation equation for the kinetic energy. Express the kinetic energy of the electron as detected from I as[7]

$$W = \frac{E_0}{\sqrt{1-\frac{u^2}{c^2}}}\left(1-\sqrt{1-\frac{u^2}{c^2}}\right) \qquad (8)$$

and the kinetic energy of the same electron as detected from I' as

$$W' = \frac{E_0}{\sqrt{1-\frac{u'^2}{c^2}}}\left(1-\sqrt{1-\frac{u'^2}{c^2}}\right). \qquad (9)$$

Dividing (8) and (9) side by side and expressing the right side of the expression obtained that way as a function of *u'* via (7) the result is

$$\frac{W}{W'} = \frac{\gamma_V(1+\beta_{u'}\beta_V\cos\theta')-\gamma_{u'}^{-1}}{1-\gamma_{u'}^{-1}} \qquad (10)$$

using the consecrated short hand notations:

$$\beta_V = V/c; \beta_{u'} = u'/c; \gamma_V = (1-V^2/c^2)^{-1/2}; \gamma_{u'}^{-1} = (1-u'^2/c^2).$$

We present in **Figure 1** the variation of *W/W'* in (10) as a function of *θ'* for the same value of *β_v*=0.5 and for different values of *β_{u'}* as a parameter whereas in **Figure 2** we present the variation of *u/u'* in (7) as a function of $\theta'$ for $\beta_V=0.5$ and different values of $\beta_{u'}$ as a parameter. The kinetic energy of a particle being an increasing function of the speed we see that the variation of *D/D'* mimics the variation of *u/u'*.

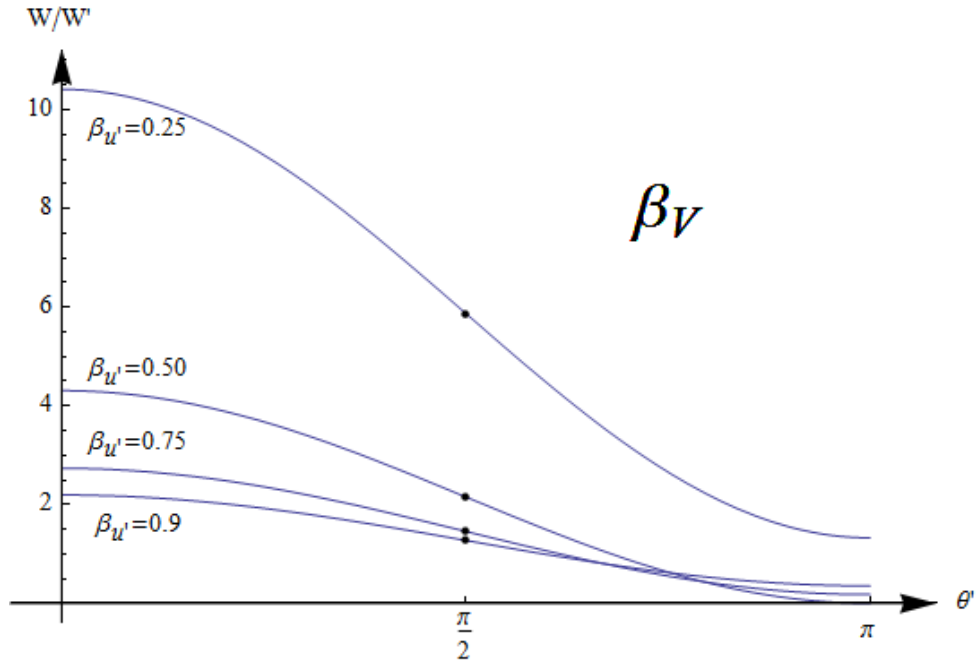

**Figure 1.** The variation of *W/W'* with the angle $\theta'$ for a constant value of $\beta_V$ and different values of $\beta_{u'}$ as a parameter.

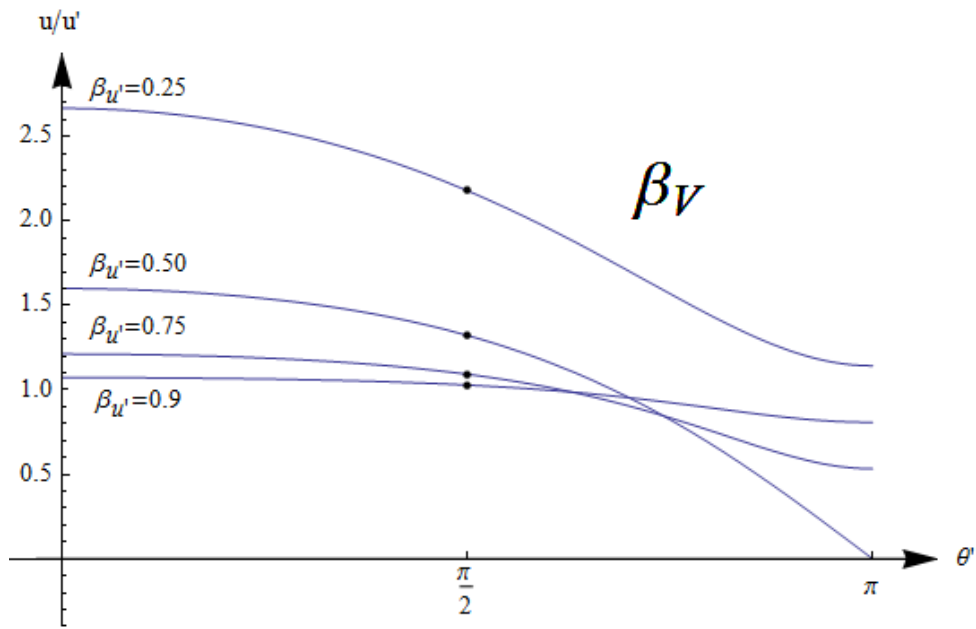

**Figure 2.** The variation of $u/u'$ with the angle $\theta'$ for a constant value of $\beta_V$ and different values of $\beta_{u'}$ as a parameter.

Applying (10) to the case of a photon, we should consider that $u'/c = 1$ and the final result being

$$W_{ph} = W'_{ph} \frac{1 + \beta_V \cos\theta'_c}{\sqrt{1 - \frac{V^2}{c^2}}} \tag{11}$$

$W_{ph}, W'_{ph}$ representing the energy of the photon as detected from I and I' respectively, $\theta_c$ and $\theta'_c$ standing for the angles made by the direction in which the photon moves with the positive direction of the overlapped axes as detected from the two involved inertial reference frames. We recognize in the right side of (11) the presence of the Doppler factor $D$

$$D = \gamma_V (1 + \beta_V \cos\theta'). \tag{12}$$

The inverse transformation of (11) reads

$$W'_{ph} = W_{ph} \gamma_V (1 - \beta_V \cos\theta_c). \tag{13}$$

Eliminating $W_{ph}$ and $W'_{ph}$ between (12) and (13) we recover the formula which accounts for the aberration of light effect

$$\cos\theta_c = \frac{\cos\theta'_c + \beta_V}{1 + \beta_V \cos\theta'_c}. \tag{14}$$

a formula which accounts for the aberration of light effect illustrating the relative character of the direction along which a photon moves.

**Figure 3** illustrates the variation of the Doppler factor $D$ given by (12) with $\theta'$ for different values of $\beta_V$. As we see for $\theta'=0$ and $\theta'=\pi$ we recover the formulas which account for the longitudinal Doppler Effect.

Starting with the definition of the kinetic energy as a difference between the energy and the rest energy

$$W = E - E_0 \tag{15}$$

and applying it to the case of a photon ($E_0=0$) we obtain

$$W_c = E_c \tag{16}$$

resulting that the energy carried by a photon or by an electromagnetic wave is kinetic energy.

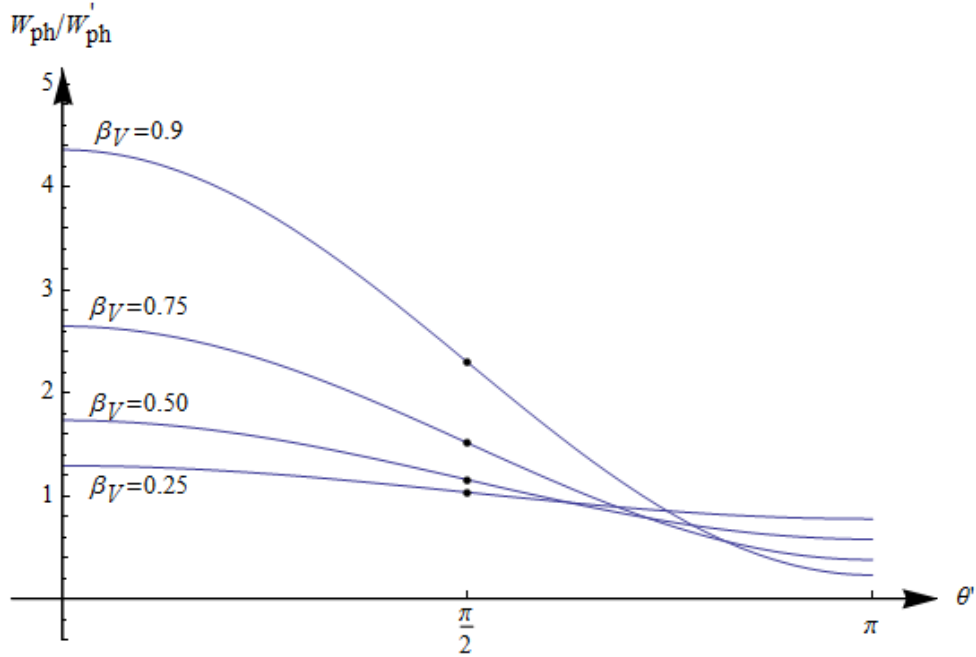

**Figure 3.** The variation of the Doppler factor $D$ (12) with the angle $\theta'$ for different values of the relative velocity $\beta_V$.

## Conclusions

The study of the direction dependence of the kinetic energy of a tardyon reveals some peculiarities of its behavior, not mentioned in the literature of the subject. As expected in all studied cases the kinetic energy is positive and in some cases there is a direction $\theta'$ along which no relativistic effects are detected. Some similarities between the behavior of the tardyon and of the photon are revealed bringing more insight in the way in which special relativity theory reflects the transition from one to the other.